\documentclass{biophys-new2}
\usepackage[utf8]{inputenc}
\usepackage{graphicx}
\usepackage[colorlinks,allcolors=cyan!70!black]{hyperref}

\usepackage{lipsum}
\usepackage{siunitx}

\papertype{Article}

\title{Evaluation of the duty ratio of bacterial flagellar motor by a dynamic load control}

\runningtitle{Evaluation of the duty ratio}

\author[1]{Kento Sato}
\author[1]{ Shuichi Nakamura}
\author[1]{ Seishi Kudo}
\author[1,*]{Shoichi Toyabe}

\runningauthor{K. Sato et al.}

\affil[1]{Department of Applied Physics, Graduate School of Engineering, Tohoku University, Sendai 980-8579, Japan}

\corrauthor[*]{toyabe@tohoku.ac.jp}

\begin{document}

\begin{frontmatter}

\begin{abstract}
Bacterial flagellar motor is one of the most complex and sophisticated nano machineries in nature.
A duty ratio $D$  is a fraction of time that the stator and the rotor interact and is a fundamental property to characterize  the motor but remains to be determined.
It is known that the stator units of the motor bind to and dissociate from the motor dynamically to control the motor torque depending on the load on the motor.
At low load where the kinetics such as a proton translocation speed limits the rotation rate,  the dependency of the rotation rate on the number of stator units $N$ infers $D$; the dependency becomes larger for smaller $D$.
Contradicting observations supporting both the small and large $D$  have been reported.
A dilemma is that it is difficult to explore a broad range of $N$ at low load because the stator units easily dissociate, and $N$ is limited to one or two at vanishing load.
Here, we develop an electrorotation method to dynamically control the load on the flagellar motor of  {\it Salmonella} with a calibrated magnitude of the torque.
By instantly reducing the load for keeping $N$ high, we observed that the speed at low load depends on $N$, implying a small duty ratio.
We recovered the torque-speed curves of individual motors and evaluated the duty ratio to be $0.14 \pm 0.04$ from the correlation between the torque at high load and the rotation rate at low load.
\end{abstract}

\end{frontmatter}



\section{Introduction}

Bacterial flagellar motor is a large complex of proteins with a dimension of about 50 nm  \cite{Berg2003,BerryArmitage1999,SowaBerry2008}.
The motor rotates a flagellar filament longer than 5 \si{\micro m} at several hundred Hz unidirectionally and propels the cell body.
For a tactic run-and-tumble response, the motor can reverse the rotation with a frequency controlled by an intracellular signal from the sensory protein on the membrane.

The motor implements an automatic torque control.
The stator of the bacterial flagellar motor is not static contrary to the name but has a dynamic structure \cite{Block1984} (Fig. \ref{Fig:intro}a).
The number of the stator units ($N$) changes depending on ion motive force (IMF) \cite{Tipping2013}, coupling ion \cite{Fukuoka2009}, and load \cite{Tipping2013, Lele2013}.
For example, when a load on the motor increased, more stator units are incorporated into the motor, and $N$ increases.

The duty ratio $D$ is a fraction of time that the stator and the rotor interact and is a fundamental property to characterize  the motor.
$D$ is inferred from the dependency of the rotation rate, $\omega$, on $N$ under a low-load condition  (Fig. \ref{Fig:intro}b).
At high load,  $\omega$  depends on $N$ independent of $D$ \cite{Ryu2000,Lo2013, Nord2017} .
This is because  the torque generation against load is the rate-limiting step at high load.
The motor generates larger torque with more stator units, resulting in a faster rotation.
On the other hand, under a low-load condition,  the kinetics such as a proton translocation speed limits $\omega$.
If $D$ is low, $\omega$ is expected to increase with $N$ because of an increased chance of the stator-rotor interaction.
If $D$ is high, $\omega$ is less affected by $N$.
However, contradicting observations were reported; some experiments implied that $\omega$ at vanishing load, referred to a zero-load speed $\omega_0$ below, is independent of $N$ \cite{Ryu2000, Yuan2008, Wang2017} and others implied that $\omega_0$ increases with $N$ \cite{Lo2013, Nord2017}.

For concluding the dependency,  it is necessary to measure $\omega_0$ in a broad range of $N$ at low load.
A dilemma is that , the stator units readily dissociate from the motor at  low load.
In previous experiments, $N$ was limited to one or two.

Here, for evaluating $D$, we used a calibrated electrorotation method to measure $\omega_0$ of a tethered {\it Salmonella enterica} serovar Typhimurium cell while keeping $N$ high.
This was achieved by quickly reducing the load on the motor by dynamic control of assisting electrorotation torque.
If the load changes sufficiently faster than  the stator dissociation dynamics, we could measure $\omega_0$ while keeping $N$ high.

We developed the calibration method of the electrorotation established previously \cite{Toyabe2010PRL,Toyabe2011PNAS, Toyabe2013Biophys, Toyabe2015NJP} for applying it to the flagellar motor of the tethered cell.
It provides not only the torque magnitude of the electrorotation torque but also the frictional coefficient felt by the motor {\it in situ} at the same time without any knowledge about the geometry of the probe-motor complex.
This implements methodology to recover the precise torque characteristics of individual motors.
Before going to the results, we briefly explain the calibration method.

\subsection{Calibrated electrorotation method}

\begin{figure}[htbp]
    \includegraphics{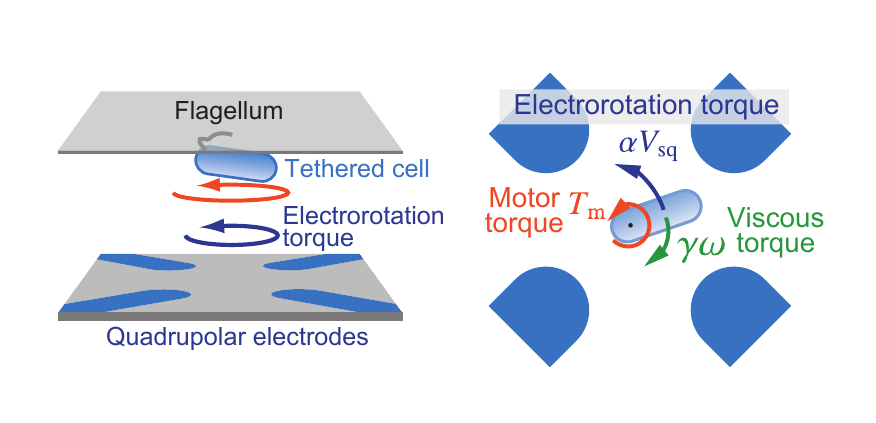} \caption{Calibrated electrorotation method imposes torque with a calibrated magnitude on a tethered cell using high-frequency electric field.
    The method provides a dynamic torque control to measure the response of the motor, which enables us to measure the torque-speed curves of individual motors.
    }
    \label{Fig:electrorotation}
\end{figure}

The electrorotation method imposes an external torque to a microscopic dielectric object by inducing high-frequency alternative current (AC) electric field.
The method has been used to apply a dynamically controlled load on the motor of tethered bacterial cells \cite{Iwazawa1993, Washizu1993, Berg1993, Berry1995, Berry1996, Berry1999, Sugiyama2004} (Fig. \ref{Fig:electrorotation}).
However, despite its potential advantages, the method is not widely used mainly due to a lack of the torque calibration method.
We previously established the calibration method and successfully applied it to an F$_1$-ATPase motor and revealed its torque characteristics \cite{Toyabe2010PRL,Toyabe2011PNAS, Toyabe2013Biophys, Toyabe2015NJP}. 
Here, we further develop the method to reduce the experimental noise and apply it to the bacterial flagellar motor.

The electrorotation torque is proportional to the square of the applied voltage amplitude $V_0$.
Therefore, the load on a motor can be dynamically and continuously controlled by modulating $V_0$.
Because inertia is negligible in microscopic systems, the torque generated by the motor $T_\mathrm{m}(\omega)$ at a rotation rate $\omega$, the viscous torque loaded on the motor,  the electrorotation torque, and the thermally fluctuating torque are balanced.
The thermal fluctuation is negligible when averaged in a sufficiently long time.
Thus, the motor torque becomes
\begin{equation}\label{eq:Tm}
T_\mathrm{m}(\omega)=\gamma\omega-\alpha V_\mathrm{sq}.
\end{equation}
Here, $\gamma$ is the rotational frictional coefficient and contains the contributions from the internal friction of the motor and the friction of the cell body and flagellar filament against the fluid.
$\alpha V_\mathrm{sq}$ is an external torque, and $V_\mathrm{sq}$ is $ V_0^2$ or $-V_0^2$ depending on the sign of the phase shift of the applied AC voltages.
The coefficient $\alpha$ depends on multiple factors including the cell shape, the chamber geometry, and the dielectric properties of the cell body and the buffer.
Therefore, it is not possible to determine $\alpha$ {\it a priori}.
This has hampered the use of the electrorotation method to measure the motor torque.
In the previous papers \cite{Washizu1993, Berg1993, Berry1995, Berry1996, Berry1999, Sugiyama2004}, ``relative'' external torque instead of the exact torque magnitude has been used to plot the torque characteristic.

%
The calibration is based on the fluctuation-response relation (FRR), which relates the thermal fluctuation to the response to external perturbation under a condition close to the equilibrium \cite{Kubo1991, ToyabeSano2015}.
Because the motor's rotation is limited to a low-frequency region and settled at equilibrium at high frequencies, the calibration is possible by comparing the rotational fluctuation and response of rotation against external perturbation at the high frequency.
%
This relation connects the fluctuation $\tilde C(f)$ and the response to a small perturbation $\tilde R'(f)$:
\begin{equation}\label{eq:FDT}
\tilde C(f)=2k_\mathrm{B}T\tilde R'(f).
\end{equation}
Here, $\tilde C(f)$ is the Fourier transform of the auto-correlation function of the rotational velocity $C(t)=\langle \omega(\tau+t)\omega(\tau)\rangle$ at a frequency $f$.
For an equilibrium Brownian movement, $\gamma$ is obtained by $\tilde C(f)=k_\mathrm{B}T/\gamma$.
On the other hand, $\tilde R'(f)$ is the real part of the frequency response of $\omega(t)$, measured by applying a small sinusoidal torque; $\tilde R(f)=w_\mathrm{r}e^{i\phi}/N_0$, where  $f$ and $N_0=\alpha V_\mathrm{sq, 0}$ are the frequency and amplitude of the sinusoidal torque. 
$w_\mathrm{r}$  and $\phi$ are the amplitude and phase of the velocity response.
By measuring the fluctuation, $\tilde C(f_0)$, and the response, $w_\mathrm{r}$ and $\phi$, of the rotation rate at sufficiently high frequency $f_0$, we can determine $\alpha_0$ and $\gamma$ using (\ref{eq:FDT});
\begin{equation}\label{eq:alpha}
\alpha=\frac{2w_\mathrm{r}\cos\phi}{V_\mathrm{sq, 0}}\gamma,\qquad \gamma=\frac{k_\mathrm{B}T}{\tilde C(f_0)}
\end{equation}

However, this calibration may be suffered by the interaction between the cell body and the glass surface, which  induces a periodic potential on the motor mechanics and yields peaks in the fluctuation spectra.
We developed a method to eliminate this effect by using an extended FRR around a local mean velocity developed by Speck and Seifert\cite{Speck2006}.
See the Materials and Methods for details.

\section{Results}

\subsection{Motor torque under low load}

\begin{figure}[htbp]
    \includegraphics{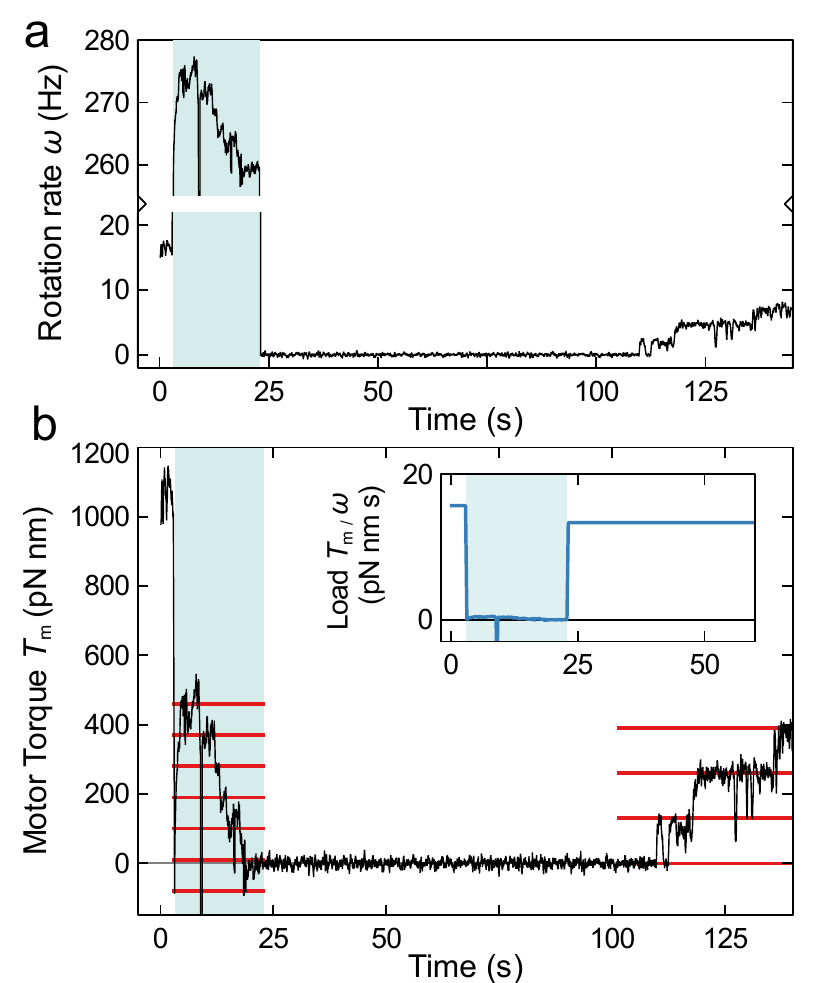}
    \caption{
    Response of the motor rotation to an assisting torque applied by the calibrated electrorotation method.
    {\bf a}, Temporal profile of the rotation rate.
    The rotation rate increased steeply by an assisting torque (indicated by cyan) and then decreased in a stepwise manner presumably due to the dissociation of the stator units.
    A steep drop of $\omega$ at around 10s might be caused by a nonspecific interaction between the cell body and the glass surface.
    {\bf b}, Motor torque $T_\mathrm{m}$ profile calculated by (\ref{eq:Tm}).
    The horizontal lines are drawn for an eye guide with a spacing of 90 pN nm (left) or 120 pN nm (right). 
        Inset: Load defined as $T_\mathrm{m}/\omega$.
    }
    \label{Fig:Resurrection}
\end{figure}

The motor of the {\it Salmonella} cell used here rotates only in the counter-clockwise direction.
For exploring the low-load region, we imposed an external torque that assists the rotation in the counter-clockwise direction.
A constant assisting torque was applied for 30s and then switched off (Fig. \ref{Fig:Resurrection}a).
The magnitude of the torque was roughly tuned so that the load on the motor is close to zero by canceling the  viscous load.
The equation (\ref{eq:Tm}) indicates that such a constant assisting torque shifts the equal-load line (Fig. \ref{Fig:intro}c, bottom).

We observed a steep rise of the rotation speed by the assisting torque and then a stepwise decrease (Fig. \ref{Fig:Resurrection}a).
The motor torque $T_\mathrm{m}$ calculated by (\ref{eq:Tm}) is shown in Fig. \ref{Fig:Resurrection}b.
Here, we used  averaged values of $\gamma$ and $\alpha$ in the periods before, during, and after the assist, respectively.
The load on the motor was quantified as $T_\mathrm{m}/\omega$, which vanished during the assist (Fig. \ref{Fig:Resurrection}b, inset).

During the assist, we observed a stepwise decrease in $\omega$ and $T_\mathrm{m}$.
After we turned off the assisting torque, the rotation stopped for a while.
Then, we observed recovery of the rotation and succeeding stepwise increase in the rotation speed.
This recovery phenomenon is supposed to be the so-called resurrection process, in which the stator units bind to the motor \cite{Block1984, Blair1988, Ryu2000, Leake2006, Fukuoka2009, Tipping2013, Lele2013}.
On the other hand, the stepwise decrease in $\omega$ at a vanishing load is thought to be the dissociation of the stator units from the motor.
These results suggest that the rotation speed at the vanishing load depends on the number of stator units $N$ (Fig. \ref{Fig:intro}c), implying a small duty ratio.
The negative torque observed during the assist is not expected in the large duty ratio (Fig. \ref{Fig:intro}c) and also supports a small duty ratio.

We observed a steep drop of  $T_\mathrm{m}$ at the beginning of the assist.
This is caused mainly by the reduction of the torque generated by each stator unit due to the load change.
This is supported by the observation that the step size of the speed during the assist is smaller than that during the resurrection process.
We do not deny the possibility that $N$ also decreased instantly at the beginning of the assist.

\subsection{Torque-speed curve}

\begin{figure}[htbp]
    \includegraphics{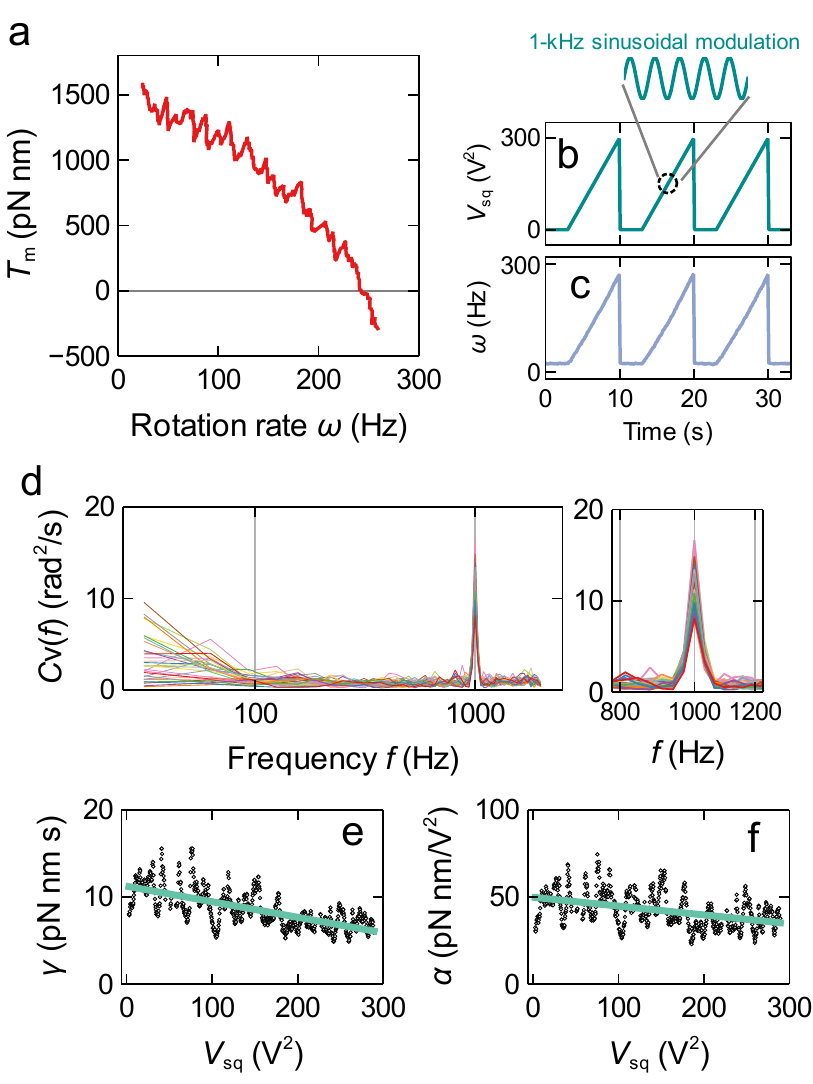}
    \caption{A typical torque-speed curve of a wild-type motor.
            {\bf a},  The motor torque $T_\mathrm{m}(\omega)$ calculated by (\ref{eq:Tm}).
    {\bf b}, Ramp cycles of $V_\mathrm{sq}(t)$. 
    We repeated a cycle of seven-second ramp from 0 to 300 $V^2$ and three-second interval.
        A ramp in a single cycle is used to recover a TS curve.
    A 1-kHz sinusoidal perturbation with the amplitude of 10 $V^2$ is superposed for the torque calibration.
    {\bf c}, Rotational rate $\omega(t)$ during cycles.
    {\bf d}, The power spectra of the rotational rate $C_\mathrm{v}(f)$ for 512-frame windows. 
    A steep peak (magnified in the right panel) corresponds to the response to the 1-kHz sinusoidal torque.
    {\bf  e} The frictional coefficient $\gamma$ calculated from  $C_\mathrm{v}(f)$ with the 512-frame windows with a 32-frame shift.
    {\bf f}, The calibration coefficient $\alpha$.
    Solid curves in {\bf e} and {\bf f} are linear-fitting curves.
    }
    \label{Fig:WT:Example}
\end{figure}

For evaluating the value of the duty ratio, we measured the torque-speed (TS) curves of the motors under an assisting torque.
We repeated a cycle consisting of a seven-second ramp from 0 $V^2$ to 300 $V^2$ and three-second intervals alternatively (Fig. \ref{Fig:WT:Example}b and c).
The torque was modulated with a 1000-Hz sinusoidal torque with an amplitude of 10 $V^2$ for evaluating  $\gamma$ and $\alpha$ {\it in situ}, with which we can calculate the motor torque by (\ref{eq:Tm}).

A typical TS curve is shown in Fig. \ref{Fig:WT:Example}a.
The TS curve reached the  zero-load state, which is not reachable by a viscous load. 
In the most experiments, $\gamma$ was smaller at higher $V_\mathrm{sq}$ (Fig. \ref{Fig:WT:Example}e).
This is because the electric field possibly induces not only a rotational torque but also a dielectric electrophoresis force that pulls the cell apart from the glass surface towards the electrodes.
This reduces $\gamma$ because of the smaller rotation radius and less surface effect \cite{Leach2009}.
The fluctuation patterns of $\gamma$ and $\alpha$ are similar because $\gamma$ is multiplied to calculate $\alpha$ (\ref{eq:alpha}). 

For validating the experimental procedure and the analysis, we applied the method to a stator-less motor.
This mutant motor lacks the stator units MotA and MotB and does not generate torque.
However, it still has a rotor and flagellar filament and exhibits a rotational Brownian motion.
The fluctuation is not free but confined  around a certain angular position.
The confinement potential is weak, and the motor can be readily rotated under an external torque.
We found that the motor torque was zero within the error (Fig.  \ref{Fig:WT:Torque-speed curves}a), validating our method.
The standard deviations were typically less than 100 pN nm/rad in a broad range of $\omega$, which indicates the accuracy of the method.
The deviation was larger for higher external torque.
This is because, with higher external torque, $T_\mathrm{m}$ is obtained by subtracting a large value of $\alpha V_\mathrm{sq}$ from a large value of $\gamma\omega$  (\ref{eq:Tm}), causing a large statistical error.
A small bump was observed around  $\omega\simeq 50$ Hz.
This was caused by the above-described confinement potential specific to the stator-less mutant, and does not affect the torque measurement of the  wild-type motors with stator units.

\subsection{Evaluation of duty ratio}

\begin{figure}[htbp]
    \includegraphics{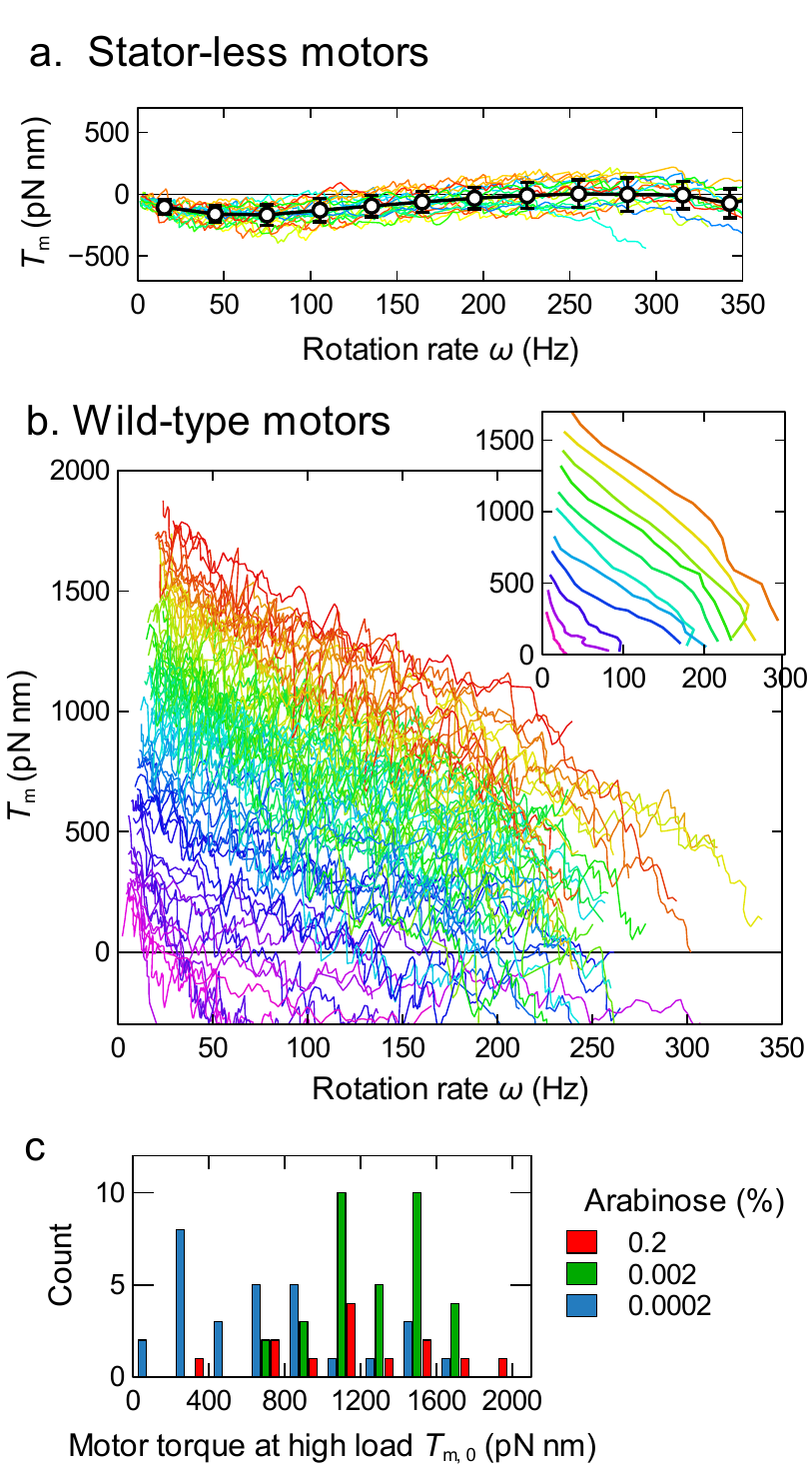}
    \caption{Torque-speed curves.
      {\bf a}, Torque-speed curves of stator-less ($\Delta$MotA/B) motors.
    The TS curves of 19 cycles of 8 motors are shown.
    The black circle is the average in the windows of 30 Hz. 
    The error bars indicate the standard deviation.
    {\bf b}, 76 TS curves of 46 wild-type motors were superposed.
    The color indicates the torque at high load. Red color indicates a high torque.
    Inset: Averaged TS curves in groups divided according to the torque at high load $T_\mathrm{m, 0}$.
     $T_\mathrm{m, 0}$ is calculated as the average of $T_\mathrm{m}$ in the first 0.4s of the ramp.
    The torque was divided in windows of 120 pN nm.
    {\bf c}, The histogram of $T_\mathrm{m, 0}$ are plotted separately for different Arabinose concentrations,  which control the expression level of the stator units.
    }
\label{Fig:WT:Torque-speed curves}
\end{figure}

For evaluating the duty ratio, we compared the TS curves of multiple motors.
In Fig. \ref{Fig:WT:Torque-speed curves}b, 76 TS curves of 46 wild-type motors are superposed.
We used cells with different expression levels of the stator units for sampling a broad range of $N$ by controlling an Arabinose concentration (Fig. \ref{Fig:WT:Torque-speed curves}c, see Materials and Methods).
We observed a broad distribution of the motor torque at high load $T_\mathrm{m, 0}$  calculated by averaging $T_\mathrm{m}$ in the first 0.4s of the ramp and a tendency that $T_\mathrm{m, 0}$ becomes smaller with a lower Arabinose concentration.
This means that we could sample a broad range of $N$ because $T_\mathrm{m, 0}\propto N$ is expected \cite{Ryu2000}.

The ramp duration of 7 s limits the duration under low load to a few seconds, which is shorter than the dynamics of the stator-unit dissociation (Fig .\ref{Fig:Resurrection}).
We expect that the dissociation of the stator units during the ramp is significantly suppressed.
However, we still observed a sudden speed change during the ramp for some curves, implying a binding or dissociation of the stator units.
For selecting the TS curves, in which $N$ is constant during the ramp, we chose the TS curves with only a slight change in the rotation speed during the ramp.
The criterion is that the change in the rotation speed during one second before and after the ramp is less than 1.5 Hz.
The TS curves containing a distinct up-and-down change, which is supposedly caused by a series of binding and dissociation of a stator unit, were also excluded from the analysis.

The curve shapes were similar among the TS curves, producing a broad distribution of the zero-load speed $\omega_0$, supporting a small duty ratio $D$.
For characterizing the TS curves, we divided them into groups according to the value of $T_\mathrm{m, 0}$ (Fig. \ref{Fig:WT:Torque-speed curves}b, inset) .
Note that we did not specify $N$ in this paper, and therefore there is no one-to-one correspondence between the group and $N$.
We found that $\omega_0$ averaged in each group had significantly distinct values.

\begin{figure}[htbp]
    \includegraphics{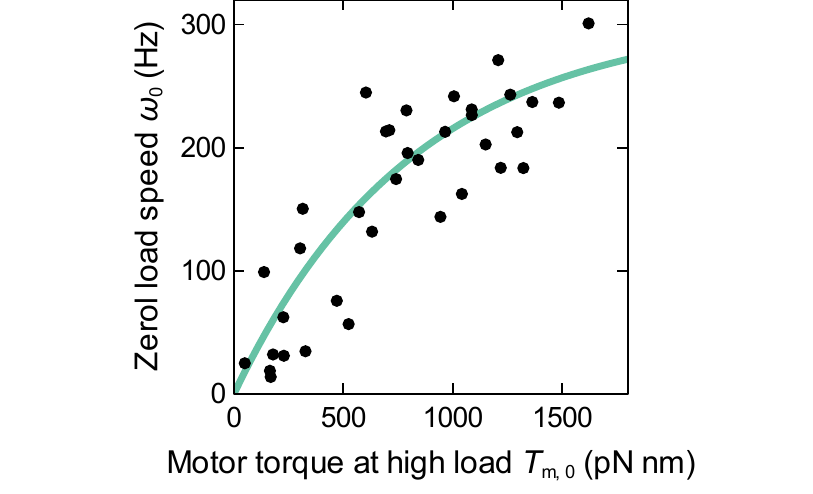}
    \caption{ The correlation between the zero-load speed $\omega_0$ and the motor torque at high load $T_\mathrm{m, 0}$.
         See Fig. \ref{Fig:WT:Torque-speed curves}b for the definition of $T_\mathrm{m, 0}$.
   A theoretical curve $\omega_0=\omega_\mathrm{0, max}\left[1-(1-D)^{T_\mathrm{m, 0}/s}\right]$ was fitted (solid curve) with fitting parameters, the maximum speed $\omega_\mathrm{0, max}$ and the duty ratio  $D$ \cite{Wang2017}.
   $s$ is a torque generated by a single stator at high load.
   We roughly estimated $s$ from the resurrection trace (Fig. \ref{Fig:Resurrection}b).
   For $s=130$ pN nm, the fitted values are $\omega_\mathrm{0, max} = 306\pm 54$ Hz and $D = 0.14\pm 0.04$. 
    }
\label{Fig:WT:Zero-load speed}
\end{figure}

We see a clear positive correlation between $\omega_0$ and  $T_\mathrm{m, 0}$  (Fig. \ref{Fig:WT:Zero-load speed}).
A simple model neglecting the interactions between the stators \cite{Wang2017} predicts a relation
\begin{equation}\label{eq:duty ratio}
 \omega_0=\omega_\mathrm{0, max}\left[1-(1-D)^{N}\right].
\end{equation}
$\omega_\mathrm{0, max}$ is the maximum rotation rate at vanishing load when $N$ is sufficiently large.
We evaluated $D$ based on (\ref{eq:duty ratio}) assuming a proportional relation $T_\mathrm{m, 0}=Ns$.
The proportional coefficient $s$ corresponds to a torque generated by a single stator at high load.
A rough estimation from the resurrection trace (Fig. \ref{Fig:Resurrection}b) gave $s=130$ pN nm.
By fitting (\ref{eq:duty ratio})  with fitting parameters  $D$ and  $\omega_\mathrm{0, max}$, we obtained $D=0.14\pm 0.04$ and $\omega_\mathrm{0, max}=306\pm 54$ Hz.
That is, each stator unit interacts with the rotor for a time fraction of 14\%.
The fitted value of $D$ varies to $0.11\pm 0.03$ and $0.16\pm 0.05$ for $s = $ 100 and 160 pN nm, respectively.

\section{Discussion and Conclusion}

Contradicting observations have been reported for the dependency of the zero-load speed $\omega_0$ on the number of the stator units $N$ at a vanishing load.
Early experiments with the H$^+$-motor of {\it Escherichia coli} implied that $\omega_0$ does not depend on $N$ \cite{Ryu2000, Yuan2008}, suggesting a high duty ratio close to one \cite{Xiang2006, Meacci2009}.
Although they did not assume that $N$ decreases at the zero-load region\cite{Tipping2013, Lele2013}, the independency of $\omega_0$ on $N$ was supported by a recent experiment with {\it E. coli}  motor \cite{Wang2017}.
On the other hand, other group reported that $\omega_0$ of chimeric Na$^+$-motor and also H$^+$-motor of {\it E. coli} depends on $N$  \cite{Lo2013, Nord2017}.
A theoretical model based on a small duty ratio of each motor explains this result \cite{Nirody2016}.
An experimental challenge regarding this apparent contradiction is that $N$ is limited to one or two under a vanishing load, and it is difficult to observe the $N$ dependency of the rotation rate.

We directly observed a stepwise reduction of the rotation rate at the low-load region, which supposedly corresponds to a change in the stator-unit number $N$.
This is an evidence to support that  $\omega_0$ depends on $N$, suggesting a low duty ratio for the {\it Salmonella} flagellar motor.

By ramping external torque at a constant rate, we recovered the TS curves of individual motors (Fig. \ref{Fig:WT:Example}).
The TS curves reached the zero-load regions, which have been estimated only by extrapolation in the methods using a viscous load such as the beads assay \cite{Chen2000, Ryu2000, Sowa2003, SowaBerry2008,Che2008,Castillo2013, Nakamura2013}.
We observed that the rotation rate does not converge to a single value in the absence of the load but had  a variation (Fig. \ref{Fig:WT:Torque-speed curves}b).
The duty ratio was determined to be $0.14\pm0.04$ from the correlation between the motor torque under a high load and $\omega_0$ (Fig. \ref{Fig:Resurrection}).
This defines how we model the rotation mechanism of the flagellar motor.
A small duty ratio might be effective for a high-speed movement, while a high duty ratio might be effective for a large-torque generation and a long processivity.
It would be intriguing to evaluates the duty ratio of the motor of different species living in different environments.

These results were obtained by fully taking the advantage of the electrorotation method for observing the dynamic response to an instant load change.
A precise measurement of the TS characteristics and dynamic load control are central to elucidate the torque generation mechanism.
A precise shape of the TS curve infers, for example, whether the torque is generated by a power-stroke type or a ratchet type mechanism \cite{Nirody2017}.
On the other hand, the measurement of the motor response to a dynamic load modulation would become a powerful tool to investigate the dynamic stator assembly.
Beads assay has been used successfully to reveal the motor properties.
It is simple and effective to know the average behavior of the motor under each condition.
However, the TS curve is blurred by averaging the torque characteristics of multiple motors in the beads assay.
Also, the method is not convenient for a dynamical response measurement, whereas there are some attempts to control the load dynamically by  changing the viscosity of a buffer  \cite{Chen2000, Inoue2008,  Che2014} or attaching a probe to the flagellum during rotation \cite{Lele2013, Wang2017}.

A future study specifying $N$ of each TS curve would be helpful to scrutinize the stator dynamics.
This may be possible by a combination of fluorescently-labeled stator units and the calibrated electrorotation.
Note that the geometry of the tethered cell has an advantage for a total internal reflection fluorescent microscopy because of the short distance between the motor and the bottom glass surface \cite{Leake2006}.
Thus, the methodology demonstrated here would add a new dimension to the study of bacterial flagellar motors.
    
Finally, we mention that the estimation of $\gamma$ based on the fluctuation does not require a pre-knowledge about the geometry of the system, and can be applied to broad systems including the beads assay of the bacterial flagellar motor, though precise noise-free fluctuation spectrum is necessary for this.
This method should be helpful to provide more reliable values of motor torque and understand the torque-generating mechanism.

This work was supported by JSPS KAKENHI (16H00791 and 18H05427).

\section{Materials and Methods}

\subsection{Electrorotation Method}

{\it Salmonella} strain YSC2123, which lacks {\it motA}, {\it motB}, {\it cheY}, {\it fimA}, and {\it fliC} (204--292), was transformed with a plasmid encoding wiled-type {\it motA/motB} \cite{Morimoto2010} (referred to as a wild-type) or not (stator-less mutant). 

Cells were grown in L-broth containing 100 \si{\micro g/ml} ampicillin for 5 hours at 30 \si{\degreeCelsius} with shaking; 0.0002\%, 0.002\%, or 0.2\% arabinose was added and then incubated for 30 minutes at 30 \si{\degreeCelsius} with shaking for protein expression (except the stator-less mutant experiments in Fig. \ref{Fig:WT:Torque-speed curves}a).
L-broth was prepared as described previously \cite{Minamino2003}.
After replacement of L-broth with the observation buffer (10 mM MOPS(3-Morpholinopropanesulfonic acid) and 10 mM KCl  adjusted to pH7.0 with KOH), we partially sheared the sticky flagella filaments by passing the bacterial solution through 25G needle 70 times. 

Observation chamber has quadrupolar electrodes with a spacing of 50 \si{\micro m} on the surface of the bottom glass slide.
After a 10-\si{\micro l} droplet of the cell solution was placed at the center of the electrodes, a coverslip (Matsunami, Japan) was placed with a thin both-side adhesive tape (10 \si{\micro m} thickness, Teraoka, Japan) as a spacer.
Only the top of the both-side adhesive tape was coated by a high-vacuum grease (Shin-Etsu, Japan) so that the top coverslip can be moved.
This enables us to move the cells tethered on the top glass slip into the center of the electrodes.
After 3 minutes, the solution was replaced with 20 \si{\micro l} of blocking buffer (observation buffer containing 50 mg/ml Perfect Block (MobiTec, Germany)) was flew into the chamber.
Perfect block serves as a blocking agent to suppress the interaction between the cell body and the glass surface.
Then, 40 \si{\micro l} of the observation buffer was flew into the chamber to wash free Perfect Block.

We observed the rotation of a tethered cell at a room temperature (24$^\circ$C) on a phase-contrast upright microscope (Olympus BX51WI, Japan) with a 60$\times$ objective lens (Olympus, NA=1.42),  at 4,000 Hz using a high-speed CMOS camera (Basler, Germany), high-intensity LED (623 nm, 4.8W, Thorlabs, NJ) for illumination, and a laboratory-made capturing software developed on LabVIEW 2014 (National Instruments, TX). 
The angular position of the cellular body was analysed by an algorithm based on a principal component analysis of the cell image.
This method reduces the instrumental noise compared to the centroid-based method.

A 5-MHz sinusoidal voltage with a phase shift of $\pi/2$ was induced on the four electrodes.
The signal was generated by a function generator (nf, Japan) controlled by PC and divided by $180^\circ$ phase distributors (Thamway, Japan).
They were amplified by four amplifiers (Analog Devices, MA) and loaded on the electrodes.
This generates an electric field rotating at 5 MHz in the center of the electrodes and induces a dipole moment rotating at 5 MHz on the cell body.
Since there is a phase delay between the electric field and dipole moment, the cell body is subjected to a constant torque.
The magnitude of torque is proportional to the square of the voltages' amplitude $V_0$ and the volume of the cell body.
We modulated $V_0$ by a signal generated by the multifunction board (National Instruments) equipped on PC.
The camera and amplitude signal were synchronized at a time difference less than one microsecond.

\subsection{Angular dependency}

The periodic perturbation, caused by the interaction between the cell body and the glass surface and also the imperfect voltage balance causes a periodic variation of the rotation rate with the frequencies equal to the integral multiple of the mean rotation rate.
These can produce nonequilibrium fluctuation at more than 1,000 Hz and affects the above torque calibration.
These nonequilibrium fluctuations are superposed on the equilibrium Brownian spectrum and impede the torque calibration.
We suppressed the nonequilibrium fluctuations dramatically by using a modified FRR derived by Speck and Seifert \cite{Speck2006, ToyabeSano2015}; $\tilde C_\nu(f)=2k_\mathrm{B}T\tilde R_\nu'(f)$.
Here, $\tilde C_\nu(f)$ and $\tilde R_\nu'(f)$ are the fluctuation and response function of the rotation rate around the local mean velocity defined as $\omega_\nu(t)\equiv \omega(t)-\nu(\theta(t))$.
$\nu(\theta_0)=\int^T_0 dt\,\delta(\theta(t)-\theta_0)\omega(t)$ is the local mean velocity at an angular position $\theta_0$.
$\delta(x)$ is the Dirac's delta function. 
The idea behind this is that the FRR is restored for the rotation around the local mean velocity.
The nonequilibrium rotations are embedded in the local mean velocity \cite{ToyabeSano2015}.
Therefore, we can estimate $\gamma$ and $\alpha_0$ by eliminating the noise by any periodic potential due to, for example, the interaction with the glass surface.
Throughout this work, we used $\tilde C_\nu(f)$ and $\tilde R'_\nu(f)$ instead of $\tilde C(f)$ and $\tilde R'(f)$ in (\ref{eq:alpha}) to eliminate the effect of the periodic potential for calculating $\gamma$ and $\alpha_0$.

\subsection{Torque-speed curve}
\label{sec:procedure}

\begin{figure}[htbp]
    \includegraphics{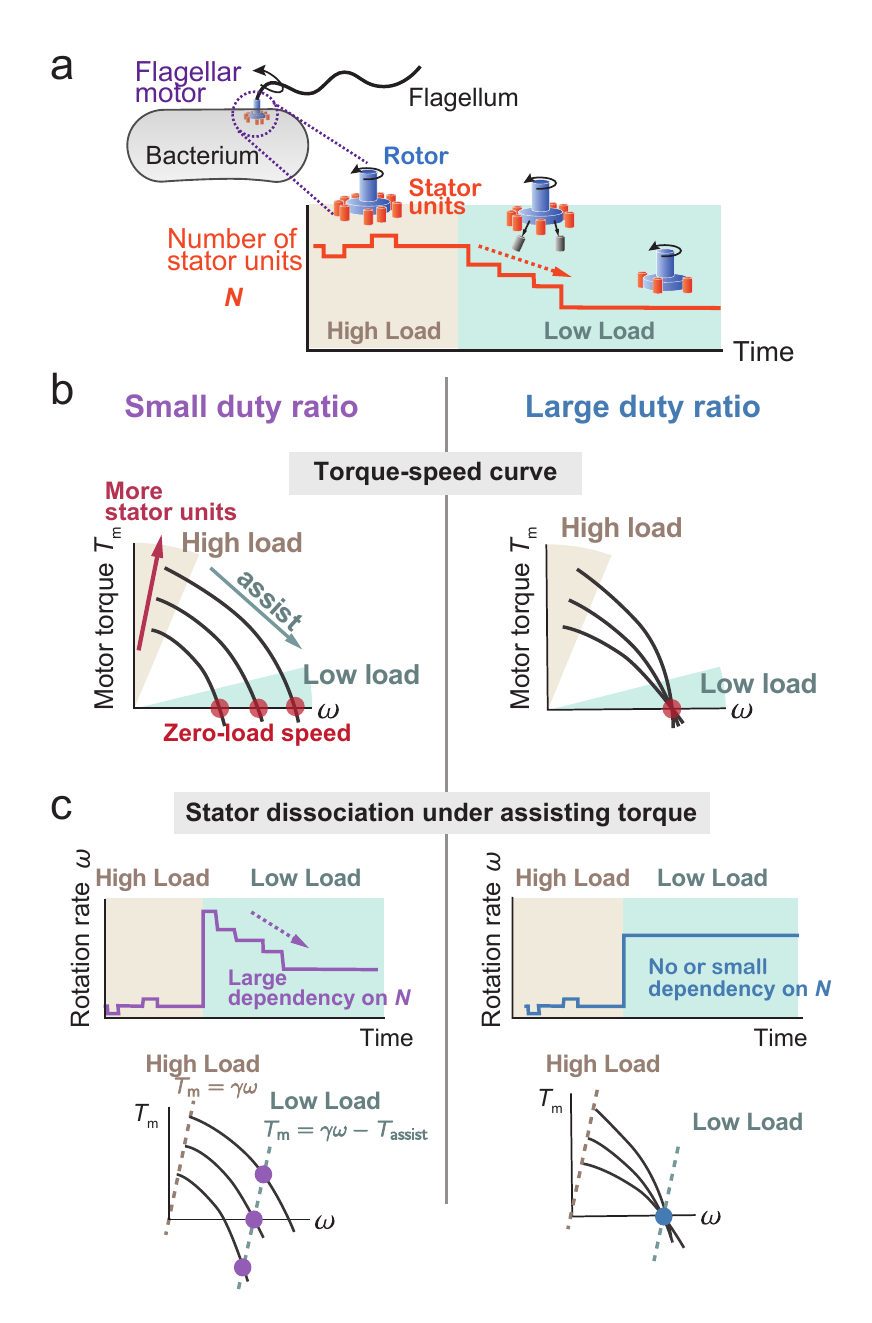}
    \caption{{\bf a}, 
The stator of the bacterium flagellar motor is a dynamic structure.
Its stator units bind to and dissociate from the motor to control the motor torque.
The number of stator units $N$ depends on the load on the motor.
{\bf b},  At high load, the rotation rate $\omega$ as well as the motor torque $T_\mathrm{m}$ changes with $N$.
On the other hand, at low load, it is not concluded if they vary with $N$.
We can use this dependency of $\omega$ and $T_\mathrm{m}$ on $N$ to evaluate the duty ratio.
Large dependency implies a small duty ratio, and small dependency implies a large duty ratio.
{\bf c},  The duty ratio is also inferred from the reversed-resurrection experiment.
As the load on the motor vanishes instantly, $N$ decreases, and then the rotation rate may decrease (small duty ratio) or may keep a similar value (large duty ratio).
The motor torque is calculated as   $T_\mathrm{m}(\omega)=\gamma\omega^2-T_\mathrm{assist}$, where $T_\mathrm{assist}$ is the external assisting torque by the electrorotation method. 
Thus, a constant assisting torque by the electrorotation translates the equal-load line parallel.
    }
    \label{Fig:intro}
\end{figure}

The experiments and analysis proceeded as follows.
We applied an external torque varied at a constant ramp rate on the tethered cell superposed by a 1000-Hz small sinusoidal torque (Fig. \ref{Fig:intro}b).
$\gamma$ and $\alpha$ had a dependency on the electrorotation strength possibly due to the pulling force towards the electric field.
Therefore, we divided the trajectory into windows with a length of 512 frames with 128-frame shifts and calculated $\gamma$ and $\alpha$ in each window.
We averaged $\tilde C_\nu(f)$ in the ranges close to 1,000 Hz: [700 Hz, 900 Hz] and [1100 Hz, 1300 Hz] to obtain $\gamma$.
$\alpha$ was calculated at 1,000 Hz by a discrete Fourier transform of the rotational trajectory.
$\gamma(t)$ and $\alpha(t)$ were smoothed by a linear fitting, and $T_\mathrm{m}$ was recovered by (\ref{eq:Tm}).

\bibliography{bj}

\begin{thebibliography}{42}
\providecommand{\url}[1]{\texttt{#1}}
\providecommand{\urlprefix}{ }

\bibitem[Berg(2003)]{Berg2003}
Berg, H.~C., 2003.
\newblock The rotary motor of bacterial flagella.
\newblock \emph{Annu Rev Biochem.} 72:19--54.

\bibitem[Berry and Armitage(1999)]{BerryArmitage1999}
Berry, R.~M., and J.~P. Armitage, 1999.
\newblock The bacterial flagella motor.
\newblock \emph{Adv Microb Physiol.} 41:291--337.

\bibitem[Sowa and Berry(2008)]{SowaBerry2008}
Sowa, Y., and R.~M. Berry, 2008.
\newblock Bacterial flagellar motor.
\newblock \emph{Q. Rev. Biophys.} 41:103--132.

\bibitem[Block and Berg(1984)]{Block1984}
Block, S.~M., and H.~C. Berg, 1984.
\newblock Successive incorporation of force-generating units in the bacterial
  rotary motor.
\newblock \emph{Nature} 309:470--2.

\bibitem[Tipping et~al.(2013)Tipping, Delalez, Lim, Berry, and
  Armitage]{Tipping2013}
Tipping, M.~J., N.~J. Delalez, R.~Lim, R.~M. Berry, and J.~P. Armitage, 2013.
\newblock Load-Dependent Assembly of the Bacterial Flagellar Motor.
\newblock \emph{mBio} 4:00551--13.

\bibitem[Fukuoka et~al.(2009)Fukuoka, Wada, Kojima, Ishijima, and
  Homma]{Fukuoka2009}
Fukuoka, H., T.~Wada, S.~Kojima, A.~Ishijima, and M.~Homma, 2009.
\newblock Sodium-dependent dynamic assembly of membrane complexes in
  sodium-driven flagellar motors.
\newblock \emph{Mol. Microbiol.} 71:825--835.

\bibitem[Lele et~al.(2013)Lele, Hosu, and Berg]{Lele2013}
Lele, P.~P., B.~G. Hosu, and H.~C. Berg, 2013.
\newblock Dynamics of mechanosensing in the bacterial flagellar motor.
\newblock \emph{PNAS} 110:11839--44.

\bibitem[Ryu et~al.(2000)Ryu, Berry, and Berg]{Ryu2000}
Ryu, W.~S., R.~M. Berry, and H.~C. Berg, 2000.
\newblock Torque-generating units of the Flagellar motor of Escherichia coli
  have a high duty ratio.
\newblock \emph{Nature} 403:444.

\bibitem[Lo et~al.(2013)Lo, Sowa, Pilizota, and Berry]{Lo2013}
Lo, C.-J., Y.~Sowa, T.~Pilizota, and R.~M. Berry, 2013.
\newblock Mechanism and kinetics of a sodium-driven bacterial flagellar motor.
\newblock \emph{PNAS} 110:E2544--E2551.

\bibitem[Nord et~al.(2017)Nord, Sowa, Steel, Lo, and Berry]{Nord2017}
Nord, A.~L., Y.~Sowa, B.~C. Steel, C.-J. Lo, and R.~M. Berry, 2017.
\newblock Speed of the bacterial flagellar motor near zero load depends on the
  number of stator units.
\newblock \emph{PNAS} 114:11603--11608.

\bibitem[Yuan and Berg(2008)]{Yuan2008}
Yuan, J., and H.~C. Berg, 2008.
\newblock Resurrection of the flagellar rotary motor near zero load.
\newblock \emph{PNAS} 105:1182.

\bibitem[Wang et~al.(2017)Wang, Zhang, , and Yuan]{Wang2017}
Wang, B., R.~Zhang, , and J.~Yuan, 2017.
\newblock Limiting (zero-load) speed of the rotary motor of Escherichia coli is
  independent of the number of torque-generating units.
\newblock \emph{PNAS} 114:12478--12482.

\bibitem[Toyabe et~al.(2010)Toyabe, Okamoto, Watanabe-Nakayama, Taketani, Kudo,
  and Muneyuki]{Toyabe2010PRL}
Toyabe, S., T.~Okamoto, T.~Watanabe-Nakayama, H.~Taketani, S.~Kudo, and
  E.~Muneyuki, 2010.
\newblock Nonequilibrium energetics of a single {F}$_1$-{ATP}ase molecule.
\newblock \emph{Phys.\ Rev.\ Lett.} 104:198103.

\bibitem[Toyabe et~al.(2011)Toyabe, Watanabe-Nakayama, Okamoto, Kudo, and
  Muneyuki]{Toyabe2011PNAS}
Toyabe, S., T.~Watanabe-Nakayama, T.~Okamoto, S.~Kudo, and E.~Muneyuki, 2011.
\newblock Thermodynamic efficiency and mechanochemical coupling of
  {F$_1$-ATPase}.
\newblock \emph{Proc. Nat. Acad. Sci. USA} 108:17951--17956.

\bibitem[Toyabe and Muneyuki(2013)]{Toyabe2013Biophys}
Toyabe, S., and E.~Muneyuki, 2013.
\newblock Experimental thermodynamics of single molecular motor.
\newblock \emph{Biophys.} 9:91--98.

\bibitem[Toyabe and Muneyuki(2015)]{Toyabe2015NJP}
Toyabe, S., and E.~Muneyuki, 2015.
\newblock Single molecule thermodynamics of ATP synthesis by {F$_1$-ATPase}.
\newblock \emph{New J. Phys.} 17:015008.

\bibitem[Iwazawa et~al.(1993)Iwazawa, Imae, , and Kobayasi]{Iwazawa1993}
Iwazawa, J., Y.~Imae, , and S.~Kobayasi, 1993.
\newblock Study of the torque of the bacterial flagellar motor using a rotating
  electric field.
\newblock \emph{Biophys. J.} 64:925--933.

\bibitem[Washizu et~al.(1993)Washizu, Kurahashi, Iochi, Kurosawa, Aizawa, Kudo,
  Magariyama, and Hotani]{Washizu1993}
Washizu, M., Y.~Kurahashi, H.~Iochi, O.~Kurosawa, S.~Aizawa, S.~Kudo,
  Y.~Magariyama, and H.~Hotani, 1993.
\newblock Dielectrophoretic measurement of bacterial motor characteristics.
\newblock \emph{IEEE Trans. Ind. Appl.} 29:286.

\bibitem[Berg and Turner(1993)]{Berg1993}
Berg, H.~C., and L.~Turner, 1993.
\newblock Torque Generated by the Flagellar Motor of Escherichia coil.
\newblock \emph{Biophys. J.} 65:2201--2216.

\bibitem[Berry et~al.(1995)Berry, Turner, and Berg]{Berry1995}
Berry, R.~M., L.~Turner, and H.~C. Berg, 1995.
\newblock Mechanical limits of bacterial flagellar motors probed by
  electrorotation.
\newblock \emph{Biophys. J.} 69:280--286.

\bibitem[Berry and Berg(1996)]{Berry1996}
Berry, R.~M., and H.~C. Berg, 1996.
\newblock Torque Generated by the Bacterial Flagellar Motor Close to Stall.
\newblock \emph{Biophys. J.} 71:3501.

\bibitem[Berry and Berg(1999)]{Berry1999}
Berry, R.~M., and H.~C. Berg, 1999.
\newblock Torque Generated by the Flagellar Motor of Escherichia coli while
  Driven Backward.
\newblock \emph{Biophys. J.} 76:580--587.

\bibitem[Sugiyama et~al.(2004)Sugiyama, Magariyama, and Kudo]{Sugiyama2004}
Sugiyama, S., Y.~Magariyama, and S.~Kudo, 2004.
\newblock Forced rotation of {Na$^+$}-driven flagellar motor in a coupling
  ion-free environment.
\newblock \emph{BBA} 1656:32.

\bibitem[Kubo et~al.(1991)Kubo, Toda, and Hashitsume]{Kubo1991}
Kubo, R., M.~Toda, and N.~Hashitsume, 1991.
\newblock Statsitical Physics II.
\newblock Springer, Berlin, second edition.

\bibitem[Toyabe and Sano(2015)]{ToyabeSano2015}
Toyabe, S., and M.~Sano, 2015.
\newblock Nonequilibrium Fluctuations in Biological Strands, Machines, and
  Cells.
\newblock \emph{J. Phys. Soc. Jpn. 84, 102001 (2015).} 84:102001.

\bibitem[Speck and Seifert(2006)]{Speck2006}
Speck, T., and U.~Seifert, 2006.
\newblock Restoring a fluctuation-dissipation theorem in a nonequilibrium
  steady state.
\newblock \emph{EPL} 74:391.

\bibitem[Blair and Berg(1988)]{Blair1988}
Blair, D.~F., and H.~C. Berg, 1988.
\newblock Restoration of torque in defective flagellar motors.
\newblock \emph{Science} 242:1678--81.

\bibitem[Leake et~al.(2006)Leake, Chandler, Wadhams, Bai, Berry, and
  Armitage]{Leake2006}
Leake, M.~C., J.~H. Chandler, G.~H. Wadhams, F.~Bai, R.~M. Berry, and J.~P.
  Armitage, 2006.
\newblock Torque-generating units of the Flagellar motor of Escherichia coli
  have a high duty ratio.
\newblock \emph{Nature} 443:355.

\bibitem[Leach et~al.(2009)Leach, Mushfique, Keen, Di~Leonardo, Ruocco, Cooper,
  and Padgett]{Leach2009}
Leach, J., H.~Mushfique, S.~Keen, R.~Di~Leonardo, G.~Ruocco, J.~M. Cooper, and
  M.~J. Padgett, 2009.
\newblock Comparison of Fax\'en's correction for a microsphere translating or
  rotating near a surface.
\newblock \emph{Phys. Rev. E} 79:026301.

\bibitem[Xing et~al.(2006)Xing, Bai, Berry, , and Oster]{Xiang2006}
Xing, J., F.~Bai, R.~Berry, , and G.~Oster, 2006.
\newblock Torque–speed relationship of the bacterial flagellar motor.
\newblock \emph{PNAS} 103:1260--1265.

\bibitem[Meacci and Tu(2009)]{Meacci2009}
Meacci, G., and Y.~Tu, 2009.
\newblock Dynamics of the bacterial flagellar motor with multiple stators.
\newblock \emph{PNAS} 106:3746--3751.

\bibitem[Nirody et~al.(2016)Nirody, Berry, and Oster]{Nirody2016}
Nirody, J.~A., R.~M. Berry, and G.~Oster, 2016.
\newblock The Limiting Speed of the Bacterial Flagellar Motor.
\newblock \emph{Biophys. J.} 111:557--564.

\bibitem[Chen and Berg(2000)]{Chen2000}
Chen, X., and H.~C. Berg, 2000.
\newblock Torque-Speed Relationship of the Flagellar Rotary Motor of {\it
  Escherichia coli}.
\newblock \emph{Biophys j.} 78:1036--41.

\bibitem[Sowa et~al.(2003)Sowa, Hotta, Homma, and Ishijima]{Sowa2003}
Sowa, Y., H.~Hotta, M.~Homma, and A.~Ishijima, 2003.
\newblock Torque-speed Relationship of the {Na$^+$-driven} Flagellar Motor of
  {\it Vibrio alginolyticus}.
\newblock \emph{J. Mol. Biol.} 327:1043--51.

\bibitem[Che et~al.(2008)Che, Nakamura, Kojima, Kami-ike, Namba, and
  Minamino]{Che2008}
Che, Y.-S., S.~Nakamura, S.~Kojima, N.~Kami-ike, K.~Namba, and T.~Minamino,
  2008.
\newblock Suppressor Analysis of the {MotB(D33E)} Mutation To Probe Bacterial
  Flagellar Motor Dynamics Coupled with Proton Translocation.
\newblock \emph{J. Bact.} 190:6660--7.

\bibitem[Castillo et~al.(2013)Castillo, Nakamura, Morimoto, Che, Kami-ike,
  Kudo, Minamino, and Namba]{Castillo2013}
Castillo, D.~J., S.~Nakamura, Y.~V. Morimoto, Y.-S. Che, N.~Kami-ike, S.~Kudo,
  T.~Minamino, and K.~Namba, 2013.
\newblock The {C}-terminal periplasmic domain of {MotB} is responsible for
  load-dependent control of the number of stators of the bacterial flagellar
  motors.
\newblock \emph{biophysics} 9:173--181.

\bibitem[Nakamura et~al.(2013)Nakamura, Kami-ike, Yokota, Kudo, Minamino, and
  Namba]{Nakamura2013}
Nakamura, S., N.~Kami-ike, P.~J. Yokota, S.~Kudo, T.~Minamino, and K.~Namba,
  2013.
\newblock Effect of intracellular {pH} on the torque-speed relationship of
  bacterial proton-driven flagellar motor.
\newblock \emph{J. Mol. Biol.} 386:332--338.

\bibitem[Nirody et~al.(2017)Nirody, Sun, and Lo]{Nirody2017}
Nirody, J.~A., Y.-R. Sun, and C.-J. Lo, 2017.
\newblock The biophysicist's guide to the bacterial flagellar motor.
\newblock \emph{Adv. Phys. X} 2:324--343.

\bibitem[Inoue et~al.(2008)Inoue, Lo, Fukuoka, Takahashi, Sowa, Pilizota,
  Wadhams, Homma, Berry, and Ishijima]{Inoue2008}
Inoue, Y., C.-J. Lo, H.~Fukuoka, H.~Takahashi, Y.~Sowa, T.~Pilizota, G.~H.
  Wadhams, M.~Homma, R.~M. Berry, and A.~Ishijima, 2008.
\newblock Torque-Speed Relationships of {Na$^+$}-driven Chimeric Flagellar
  Motors in {Escherichia coli}.
\newblock \emph{Biophys. J.} 376:1251--1259.

\bibitem[Che et~al.(2014)Che, Nakamura, Morimoto, Kami-ike, Namba, and
  Minamino]{Che2014}
Che, Y.-S., S.~Nakamura, Y.~V. Morimoto, N.~Kami-ike, K.~Namba, and
  T.~Minamino, 2014.
\newblock Load-sensitive coupling of proton translocation and torque generation
  in the bacterial flagellar motor.
\newblock \emph{Mol. Biol.} 91:175--184.

\bibitem[Morimoto et~al.(2010)Morimoto, Che, Minamino, and Namba]{Morimoto2010}
Morimoto, Y.~V., Y.~S. Che, T.~Minamino, and K.~Namba, 2010.
\newblock Proton-conductivity asssay of plugged and unplugged {MotA/B} proton
  channel by cytoplasminc {pHluorin} expressed in {Salmonella}.
\newblock \emph{FEBS Lett.} 584:1268--72.

\bibitem[Minamino et~al.(2003)Minamino, Imae, Oosawa, Kobayashi, and
  Oosawa]{Minamino2003}
Minamino, T., Y.~Imae, F.~Oosawa, Y.~Kobayashi, and K.~Oosawa, 2003.
\newblock Effect of intracellular {pH} on rotational speed of bacterial
  flagellar motors.
\newblock \emph{J. Bac.} 185:1190--1194.

\end{thebibliography}

\clearpage

\end{document}